\documentclass{PoS}
\usepackage{xspace}
\usepackage{multicol}

\newcommand{\apenetp}{APEnet+\xspace}
\newcommand{\apelink}{APElink\xspace}
\newcommand{\apelinkp}{APElink+\xspace}
\newcommand{\ie}{\textit{i.e.}\xspace}
\newcommand{\eg}{\textit{e.g.}\xspace}
\newcommand{\etc}{\textit{etc.}\xspace}

\title{\apenetp: a 3D toroidal network enabling Petaflops scale Lattice QCD simulations on commodity clusters}

\ShortTitle{\apenetp}

\author{\speaker{Roberto Ammendola}$^{ab}$, Andrea Biagioni$^c$, Ottorino Frezza$^c$, Francesca Lo Cicero$^c$, Alessandro Lonardo$^c$, 
        Pier Paolucci$^c$, Roberto Petronzio$^{ad}$, Davide
        Rossetti$^c$, Andrea Salamon$^a$, Gaetano Salina$^a$, Francesco Simula$^e$, Nazario Tantalo$^{ab}$,
        Laura Tosoratto$^c$ and Piero Vicini$^c$\\
\llap{$^a$}INFN Roma Tor Vergata\\
\llap{$^b$}Centro Studi e Ricerce e Museo della Fisica \textquotedblleft Enrico Fermi \textquotedblright\\
\llap{$^c$}INFN Roma La Sapienza\\
\llap{$^d$}Universit\`{a} di Roma Tor Vergata\\
\llap{$^e$}Universit\`{a} di Roma La Sapienza\\
E-mail: \email{roberto.ammendola@roma2.infn.it}}

\abstract{ 
  Many scientific computations need multi-node parallelism for
  matching up both space (memory) and time (speed) ever-increasing
  requirements. The use of GPUs as accelerators introduces yet another
  level of complexity for the programmer and may potentially result in
  large overheads due to the complex memory hierarchy. Additionally,
  top-notch problems may easily employ more than a Petaflops of
  sustained computing power, requiring thousands of GPUs orchestrated
  with some parallel programming model.
  
  Here we describe \apenetp, the new generation of our
  interconnect, which scales up to tens of thousands of nodes with
  linear cost, thus improving the price/performance ratio on large
  clusters.
  The project target is the development of the \apelinkp host adapter
  featuring a low latency, high bandwidth direct network,
  state-of-the-art wire speeds on the links and a PCIe X8 gen2 host
  interface. It features hardware support for the RDMA programming
  model and experimental acceleration of GPU networking.
  A Linux kernel driver, a set of low-level RDMA APIs and an OpenMPI
  library driver are available, allowing for painless porting of
  standard applications.

  Finally, we give an insight of future work and intended
  developments.

}

\FullConference{The XXVIII International Symposium on Lattice Field Theory, Lattice2010\\
		June 14-19, 2010\\
		Villasimius, Italy}

\begin{document}


\section{Introduction}

When we started our research on custom cluster interconnect back in
2003, we were at the beginning of the cluster revolution in a HPC
arena which used to be dominated by custom supercomputers.
Around that time, some seminal works~\cite{Luscher} demonstrated the
effective use of mainstream processors on Lattice QCD as well as on
many other numerical problems.
Not moving from the field of LQCD, another pioneering paper appeared
some time afterward~\cite{first_GPU_LQCD}, showing that the same
codes could be easily implemented on the newborn GPGPU architectures
--- with costs that didn't deviate from those of, say, an ordinary PC
workstation equipped with a medium/high-end video card ---, displaying
a substantial performance increase; research in this area is still
progressing~\cite{clark}.

Today, to stay on a sustainable and \textit{green} path towards the
10 Petaflops barrier, cluster systems are morphing into hybrid
systems, starting from RoadRunner to current GPU accelerated
systems.
%
The TOP10 November 2010 list, the top 10 systems among the Top500
(\texttt{www.top500.org}), enlists 3 GPU accelerated clusters and,
perhaps more significantly, 6 between custom and proprietary network
interconnected systems.
%
Today as seven years ago, we think that there is still room for
innovation in the area of custom interconnection networks.
Take for example a hypothetical 4 Petaflops GPU cluster --- similar in
performance to recently introduced GPU-based HPC systems; --- it can
be currently assembled out of 6000 GPU accelerated nodes, crammed into
roughly 90-100 compute cabinets plus additional storage and
communication cabinets.
Connecting together 6000 nodes with current top-performing Infiniband
QDR technology, \eg using a full bi-sectional bandwidth tree, means
building a multi-level fat tree of switches which easily imposes hard
engineering problems --- reliability, stability, \etc --- and
unexpectedly high costs, as high as 2000\$ per port, properly
accounting for the costs of the adapters, switches, cables both
between cards and switches and between switches at different
levels.
Surprisingly, a high-performance interconnect can amount to 30-40\% of
the cost of a computing node.

Drawing on our past experience with the design of custom hardware for
Lattice QCD HPC ~\cite{APE,apeNEXT-lat1,apeNEXT-lat2,apeNEXT},
together with further developments in EU Framework Programme 6 project
SHAPES ~\cite{SHAPES1,DNP1} we designed \apenetp, a new implementation
of our 3D torus cluster interconnect ~\cite{apenet2004,apenet2005},
based on the \apelinkp adapter, which integrates both a network
interface and a switching component~\footnote{This is commonly
  referred to as a \emph{direct network}, as there is a packet
  switching component integrated into each network node, as opposed to
  traditional \emph{indirect networks} like Ethernet and Infiniband,
  where switching capabilities are moved to dedicated hardware
  components.}.
The advantages of a torus network over a switched fabric network ---
\ie Infiniband --- are clear and well known: it naturally suits the
transmission patterns of a broad range of numerical simulation codes
that are parallelized using the domain decomposition approach (which
is the case with our codes for LQCD). It is performant, its
bi-sectional bandwidth being conserved when cluster size is scaled to
large volumes. It grants linear scaling of costs without additional
expenses, the unitary cost of the \apelink adapter and three cables
being the only factor.

The downside clearly lies in the additional number of cables, with the
necessary planning for the routing of cables of assorted lengths,
according to the distance they have to travel, \ie inside or outside
each cabinet.
Additionally, a torus interconnect gradually degrades its performance
as the application communication pattern gets more irregular --- \eg
protein folding, fluid dynamics, \etc --- In these cases, an initial
pre-processing operation may be necessary to optimally map the problem
onto the underlying interconnect.



\section{The \apenetp hardware}
\label{sec:hw}
In this section, we introduce the \apenetp interconnect, our low
latency, high bandwidth direct network, supporting state-of-the-art
link wire speeds and a PCIe X8 gen2 host connection.  The \apenetp
project is based on the \apelinkp, an FPGA-based board, which is
discussed in section~\ref{apelink_card}.

On this network, the computing nodes --- \eg a multi-core CPU
optionally paired with GPU --- are arranged in a cubic mesh
interconnected by point-to-point links to form a 3D torus;
thus, each node has 6 bi-directional full-duplex communication
channels, \ie along the $X+$, $X-$, $Y+$, $Y-$, $Z+$ and $Z-$
directions.
Packets have a fixed size envelope (header+footer) and are auto-routed
to their final destinations according to wormhole dimension-ordered
static routing, with dead-lock avoidance.


\subsection{Architecture Outlook}
%
A computing host can be equipped with one such board and made into a
low latency, high bandwidth cluster node.


The \apenetp architecture may be seen as a network of routers --- see
router component in Fig.~\ref{fig:internals} --- with
configurable routing capabilities operating on packets with payload of
variable size.

The torus link block --- see top part of Fig.~\ref{fig:internals} ---
manages the data flow by encapsulating the \apenetp packets into a
light, low-level, \emph{word-stuffing} protocol able to detect transmission
errors via CRC.
It implements two virtual channels~\cite{Dally} and proper
flow-control logic on each RX link block to guarantee deadlock-free
operations.

The \apelinkp network interface --- see bottom part of
Fig.~\ref{fig:internals} --- has basically two main tasks:
\begin{itemize}
\item On the transmit data path, it gathers data coming in from the
  PCIexpress port, fragmenting the data stream into packets which are
  forwarded to the relevant destination ports, depending on the
  requested operation.
\item On the receive side, it provides hardware support for the RDMA
  programming model, implementing the basic RDMA capabilities (PUT and
  GET semantics) at the firmware level.
\end{itemize}
A micro-controller --- the NIOS II 32 bit embedded processor, which is
a standard Altera\textsuperscript{\textregistered} Intellectual
Property --- mainly helps in simplifying the implementation of the
receive path.


\subsection{The \apelinkp card}
\label{apelink_card}

The \apelinkp card is the latest generation of the \apelink hardware,
leveraging the most recent advances in host interface technology,
physical link speed and connector mechanics --- see
Table~\ref{tab:card_features}. ---
The \apelinkp card is a single FPGA-based PCI Express board,
representing a vertex of a 3D torus mesh network with 6 independent
point-to-point multiple links channel (i.e. the links between mesh
sites).

The employed FPGA device is the EP4SGX290 which is part of the
Altera\textsuperscript{\textregistered} 40 nm Stratix IV device
family.
This device is equipped with 36 full-duplex CDR-based transceivers,
each supporting data rates up to 8.5 Gbps.
Each link is made up of 4 bidirectional lanes bonded together with
proper alignment logic.
%
%

%
High-level functions, like RDMA virtual-to-physical address tables
look-up, are carried out by a program running on the FPGA embedded
micro-controller (NIOS II) which uses the DDR3 module as both program-
and data-memory.

\begin{figure}\vspace{-20pt}
\centering
  \includegraphics[width=.6\textwidth]{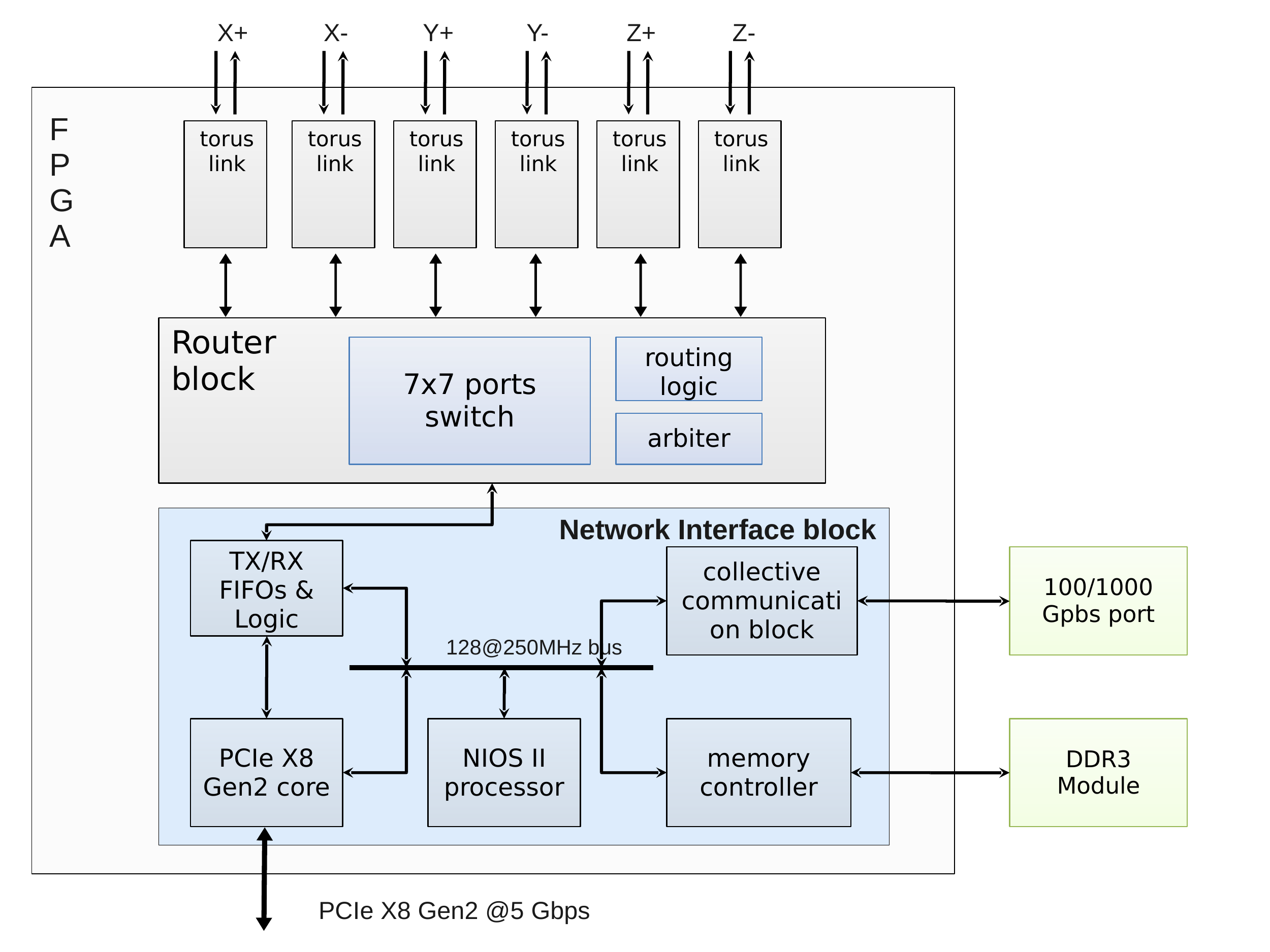}
  \caption{Internal FPGA block architecture.}
  \label{fig:internals}
\end{figure}

The hardware block structure, depicted in Fig.~\ref{fig:internals}, is
split into a so called \emph{network interface} --- the packet
injection and processing logic comprising PCIe, TX/RX logic, \etc ---
a \emph{router} component and multiple \emph{torus links}.

The torus links are 6 independent blocks with 2 virtual channel
receive buffers needed for deadlock prevention.
Proper flow control is maintained via credits handshake between a local
RX block and the remote TX block, and it is embedded in the link
protocol data layer.
The torus link is autonomously able to re-transmit the header and the
footer in case of transmission errors.
Therefore the protocol assures the delivery of the packet avoiding
nonrecoverable situations where badly corrupted packets (with errors
in the header or footer) pose a threat to the global routing.
Packets with payload errors (signaled by the footer) are handled at
the software level.
The chosen CRC polynomial generator is the industry-standard,
well-known CRC-32.

The router comprises a fully connected, 7-ports-in/7-ports-out switch, plus
routing and arbitration blocks.
The routing block examines a packet header and resolves the destination address
to a proper path across the switch.
It supports the dimension-ordered routing algorithm, with a routing
latency of 60ns.

The host interface, implemented as PCIe X8 gen2, allows communication between the
host processor and the network.

\begin{table}
\centering
\scriptsize
\begin{tabular}{|c||c|c|}
\hline
                & {\bf \apelink}    & {\bf \apelinkp}        \\
\hline
\hline
FPGA component  & Altera Stratix S30 & Altera Stratix IV GX 290 \\
\hline
\# links        & 6          & 4/6             \\
\hline
link technology & external National ser/des & embedded Altera transceivers  \\
\hline
link cables     & LVDS       & QSFP+ standard  \\
\hline
raw link speed  & 6 Gbps     & 34 Gbps         \\
\hline
host interface  & PCI-X 133MHz  & PCIe X8 Gen2  \\
\hline
peak host BW    & 1GB/s       & 4+4GB/s         \\
\hline
\end{tabular}
\normalsize
\caption{Evolution of the \apelink cards.}
\label{tab:card_features}
\end{table}



Moreover, an Ethernet port is foreseen in order to build an
additional, secondary network with an offload engine for collective
communication tasks.




\section{The \apenetp software}
\label{sec:sw}
All \apenetp software is developed and tested on RedHat Enterprise
Linux 5 and is available under the GNU GPL Licence.
It spans across four major topics: the firmware software running on
the FPGA embedded processor --- the NIOS II processor in
Fig.~\ref{fig:internals}, --- the linux kernel driver, the application
level RDMA library and a MPI implementation.
We developed a native \apenetp BTL module for OpenMPI 1.X, implemented
on top of the RDMA APIs.

The firmware software running on the FPGA embedded processor is
currently in charge of managing the RDMA virtual-to-physical address
translation table, but we are exploring new ways to exploit it for
higher-level tasks.

For maximum performance, applications can use a set of low-level
custom RDMA APIs, which is available as a C language library:
\begin{itemize}
\item Communication primitives available to applications are:
  \texttt{rmda\_put()}, \texttt{rdma\_get()}, \texttt{rdma\_send()}.
\item Memory buffer registration primitives allow for exposing memory
  buffers to RDMA primitives: \texttt{register\_buffer()},
  \texttt{unregister\_buffer()}.
\item Events are routed to applications whenever RDMA primitives are
  executed by \apenetp: \texttt{wait\_event()}.
\end{itemize}


%
We are currently working~\cite{gtc2010} on the hardware and software
features needed for GPU-initiated communications, \eg providing a
NVidia CUDA
version of the \texttt{rdma\_put()} primitive, using so called PCIe
peer-to-peer transactions, in order to avoid intermediate copies onto
CPU memory buffers.
Along the same lines of overhead reduction, there is work underway for
implementing RDMA events delivery --- by the \apelinkp hardware in CPU
memory --- accessible from within CUDA kernels.

Another research topic is exposing GPU memory areas as RDMA buffers,
in such a way they can be target of RDMA PUT and GET operations, so
cutting the latency of network operations. To this end, discussions
are ongoing with some GPU vendors.


\section{The deployment initiative}
\label{sec:quong}
We are aggregating a small community of LQCD developers and users
around our \textit{QUonG} (lattice QUantum chromodynamics ON Gpus)
deployment initiative.
\begin{figure} \vspace{-20pt}
  \centering
  \includegraphics[width=.4\textwidth]{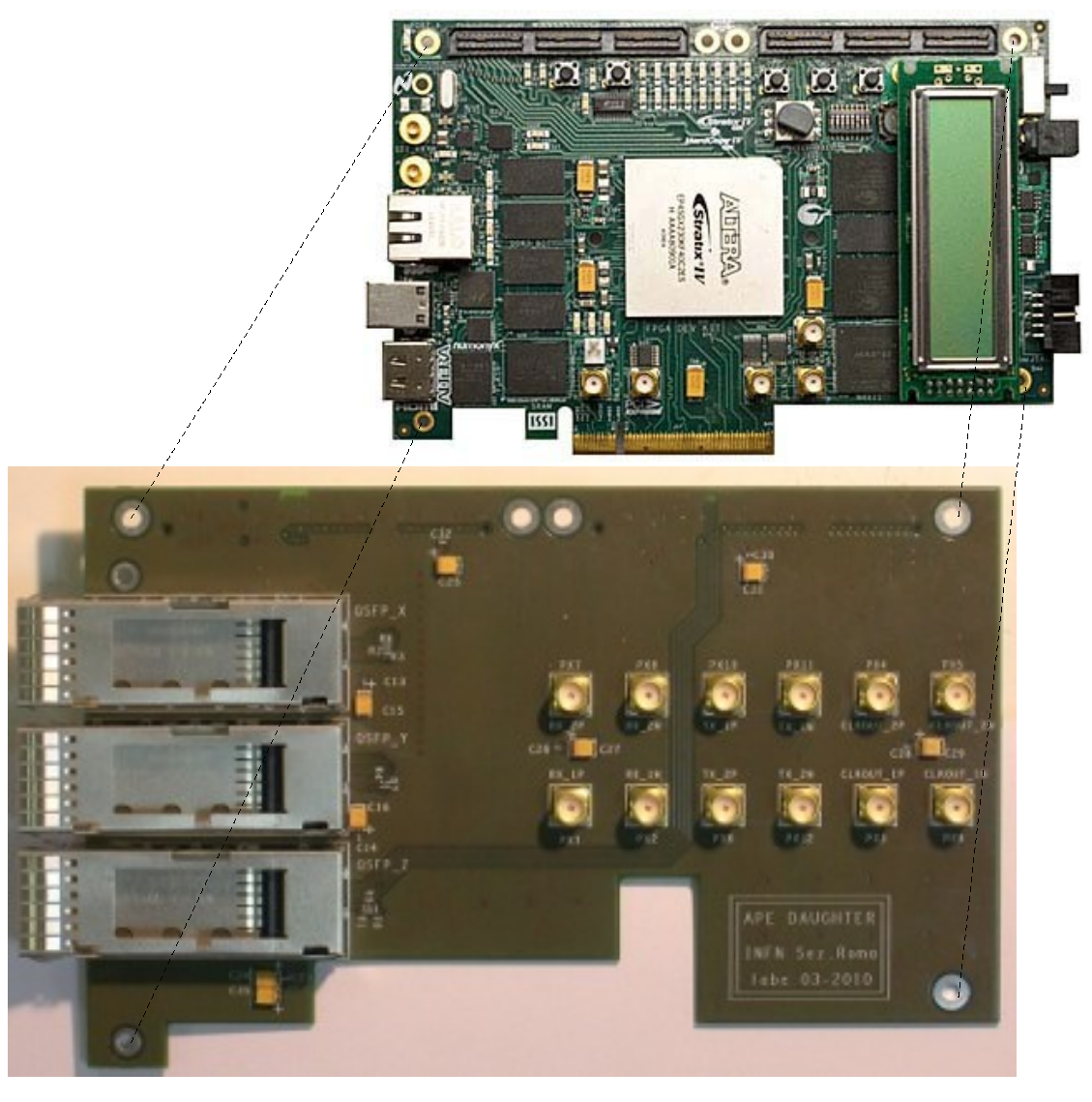}
  \hskip 2em
  \includegraphics[width=.4\textwidth]{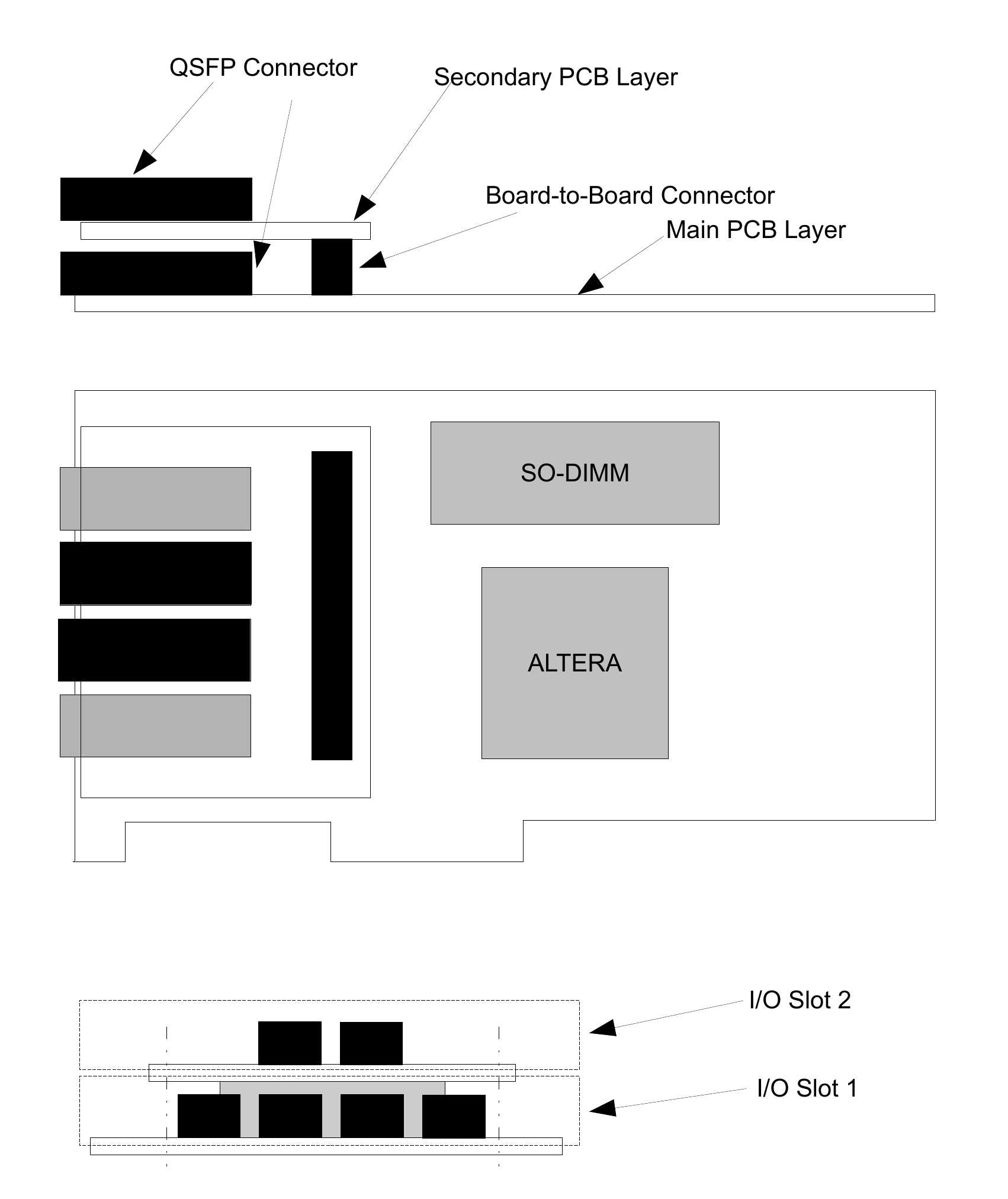}
  \caption{Altera Development Kit with our custom daughter-card (left). Schematic drawings of the \apelinkp device (right).}
  \label{fig:board}
\end{figure}
The reference platform is a GPU-accelerated cluster with the following
characteristics: a single enclosure system with at least 16 computing
nodes, \eg with a $4x2x2$ Torus topology. Each node is a dual-socket
CPU system in a 1U enclosure with at least 2 X16 PCIexpress slots.
One \apelink adapter per CPU. One or two GPUs, possibly using separate
enclosures --- \eg NVidia Tesla S2050. ---
The software environment is mainly OpenMPI or an hybrid OpenMP plus
OpenMPI.
Some optimized LQCD kernels will be provided, mainly consisting of the
GPU optimized Wilson-Dirac kernel and a simple multi-node parallel
solver.


\section{Project status and future developments}
\label{sec:status}


The \apenetp hardware is at its final stages of development. A
schematic view of the complete board is visible in
Figure~\ref{fig:board}.
Four links out of six are hosted on the main board, and two more, say
$Z+$ and $Z-$, are located in a small daughter-card on the upper
level.
In this way, the complete card occupies two PCI standard slots in a PC
chassis, while it's still possible to use it in a four-link and only
one slot wide configuration.
The prototypes will be available at beginning of 2011.


Meanwhile, a test system has been implemented in order to develop the
FPGA firmware, the PCI Express interface and the physical layer
interconnection technology. We used a commercial
Altera\textsuperscript{\textregistered} development kit (equipped with
a smaller Altera\textsuperscript{\textregistered} Stratix IV GX 230)
and a custom-designed daughter-card (an HSMC mezzanine designed at
LABE in INFN-Roma) hosting 3 QSFP+ connectors.
This assembled system allows us to test the QSFP+ technology together
with the embedded Altera\textsuperscript{\textregistered} transceivers
up to a bit rate of 24 Gbps for each link. Extensive electrical
characterization is in progress (\cite{twepp}, \cite{chep} and more
work yet to be published).


A first mini-cluster is being assembled together with GPUs and the
\apelinkp version with 3 links, for final validation of the firmware,
the interconnection and the complete software stack on a small size
(2-8 nodes).
Synthetic tests, as well as real life simulations, will be performed
by the end of 2010, so to be ready with the 6-links prototype release
and eventually a bigger cluster deployment.

We thank the people in the Electronics Laboratory (LABE) in INFN-Roma,
in particular Valerio Bocci, Giacomo Chiodi and Riccardo Lunadei.

This work was partially supported by the EU Framework Programme 7
project EURETILE under grant number 247846.




\end{document}